\documentclass[aps,12pt,floatfix]{revtex4-2}
\usepackage{amsmath}
\usepackage{graphicx}
\begin{document}
\title{Quantitative Interpretation of\\ Simulated Polymer Mean-Square Displacements}
\author{George D. J. Phillies}
\email{phillies@wpi.edu}
\affiliation{Department of Physics, Worcester Polytechnic Institute,Worcester, MA 01609}
\begin{abstract}
We propose a path for making quantitative analyses of mean-square displacement curves of polymer chains in the melt or in solution.   The approach invokes a general functional form that accurately describes $g(t) \equiv \langle (\Delta x(t))^{2} \rangle$ for all times at which $g(t)$ was measured, and that gives values for the logarithmic derivative of $g(t)$ for the same times.  By these means we can readily distinguish between regimes in which $g(t)$ follows a power law in time,  does not follow a power law in time, or has an inflection point. In a power-law regime, the method accurately determines the exponent, without imposing any assumption as to the exponent's value.
\end{abstract}
\maketitle

\section{\label{sec:1} Introduction}

Many traditional models of polymer solution dynamics predict that the mean-square displacement $g(t)$ of polymer molecules in solution or in the melt has the form of a power-law dependence\cite{degennes1979a,doiedwards1986a}
\begin{equation}
   g(t) \equiv  \langle (\Delta x(t))^{2}\rangle  = A t^{\alpha},
\label{eq:meanssqdx2}
\end{equation}
where here $\Delta x(t)$ is a time-dependent displacement, $\langle \cdot \rangle$ denotes an average, $t$ is an elapsed time, $\alpha$ is an exponent, and $A$ is a prefactor.  A power-law dependence is predicted to be valid over some time interval, multiple power laws separated by transition regimes serving to cover a full range of times.  Several displacements have been considered by different authors, including displacement of the polymer center-of-mass, displacement of individual polymer beads, or displacement of some number of polymer beads near a polymer's midpoint.  The models predict the order in which different power-law regimes and exponents will be encountered with increasing time, but do not predict quantitatively the times at which the transitions occur. The models do not in general predict the prefactor $A$, the width of the transition regions, or the time dependence of $g(t)$ in the transition regimes between the power-law regions. To display $g(t)$ as a function of time, recourse is uniformly had to log-log plots, on which power laws such as that seen in equation \ref{eq:meanssqdx2} appear as straight lines, each line's slope revealing an exponent $\alpha$.

Our objective in this proof-of-principle note is to propose a quantitative method for analyzing the time dependence of $g(t)$. The proposed method is generally applicable, in that it is model-independent.  It does not assume that $g(t)$ has a particular functional form, such as the power law of equation \ref{eq:meanssqdx2}.

Our approach is to invoke a general functional form that uniformly describes the time dependence of $g(t)$ over the entirety of times for which $g(t)$ has been obtained, without any assumption other than that the original data is consistent with a continuous line. Reported measurements appear to lie on smooth, monotonically-increasing curves, so a reasonable choice of general form is a finite Taylor series expansion.  The statistical spread in the data, in the reports we have examined, appears to be relatively independent of time on a log-log plot, so it is appropriate to take $z = \log(t)$ and write
\begin{equation}
    \label{eq:taylor}
    \log(g(t)) = \sum_{i=0}^{N} a_{i} z^{i},
\end{equation}
for the series.  Here the $a_{i}$ are the fitting parameters and $N$ is the order of the fit. The $a_{i}$ are obtained by linear-least-squares. The scatter in measurements of $\log(g(t))$ is nearly independent of $t$, so issues concerning the statistical weights to be assigned to different points do not arise. As a cautionary note, Taylor series are here used as interpolants. It is not claimed the series would be valid as an extrapolant beyond the range of $t$ of the original measurements.  What order of fit is appropriate?  By increasing $N$, one reaches values of $N$ for which the Taylor expansion agrees with the measurements, with further modest increases in $N$ having no significant effect on the form of the fitted curve.

To clarify the behavior of $g(t)$, we also determined its first and second logarithmic derivatives, the first logarithmic derivative being
\begin{equation}\label{eq:derivative}
     \frac{d \log(g(t))}{d \log(t)} = \sum_{i=1}^{N} i a_{i} z^{i-1}.
\end{equation}

Our general search for papers on simulations of polymer dynamics revealed a considerable number of reports of $g(t)$ for one or another definition of $\Delta x(t)$.  The reports are, uniformly, graphical. Measurements were digitized using Un-Scan-It 7.0, for the most part in manual mode.  Numerical analysis was made using Mathematica 12.1.

This is a proof-of-principle paper. The objective is to demonstrate that our method works, not to discuss in detail what it reveals about polymer physics.  We selected three measurements of $g(t)$ that serve to demonstrate significant aspects of our approach, namely Padding and Briels\cite{padding2001b}, their Figure 1 showing the mean-square atomic displacements, Brodeck, et al.\cite{brodeck2010a}, their Figure 1 with the 400 K center-of-mass displacements, and Peng, et al.\cite{peng2017a}, their Figure 9a with the mobility of their B beads.  Why did we chose these three data sets?  Analysis of the Padding and Briels $g(t)$ tests if the approach can usefully represent simple power-law behavior. One might be concerned as to how large an $N$ is needed to represent a $g(t)$ with multiple features, and what consequences would follow if the order of the fit was increased above the order needed to represent $g(t)$ accurately.  The effect of increasing the fit order is revealed by study of a $g(t)$ from Brodeck, et al.\cite{brodeck2010a}  Finally, around an inflection point in $g(t)$, a tangent line might mimic a power law.  We inquire if actual power law regimes and inflection points can be distinguished, using results of Peng, et al.\cite{peng2017a}.

It is legitimate to ask if the approach described here is generally applicable, and if the method leads to new physical results. A full length paper, answering these questions by analyzing close to a hundred sets of $g(t)$ data as obtained by more than a dozen research groups, is now in preparation.

\section{Tests of Finite Taylor Series as a General Functional Form}

In a series of papers, Padding and Briels\cite{padding2001b,padding2001c,padding2002a,padding2003b} report simulations of a united-atom model for linear polyethylene, reporting among other dynamic quantities the diffusion coefficient, the shear relaxation modulus, the end-to-end vector's time autocorrelation function, the single-chain coherent dynamic structure factor, and, of central interest here, various mean-square displacements. Figure 1 shows Padding and Briels' determination of $g_{\rm at}(t)$, the mean-square displacement of their individual united atoms.  We fit $\log(g(t))$ to an eighth-order polynomial in $\log(t)$. The circles in the Figure are data points extracted from the original paper.  When the points were too densely packed to be read clearly, the circles are sampled representations of the original measurements. The solid line that passes through the circles shows our best-fit polynomial.  The dashed line represents the slope, the first logarithmic derivative, of $g(t)$. One sees that the agreement between our fit and the original measurements is excellent.

\begin{figure}[thb]
  \includegraphics{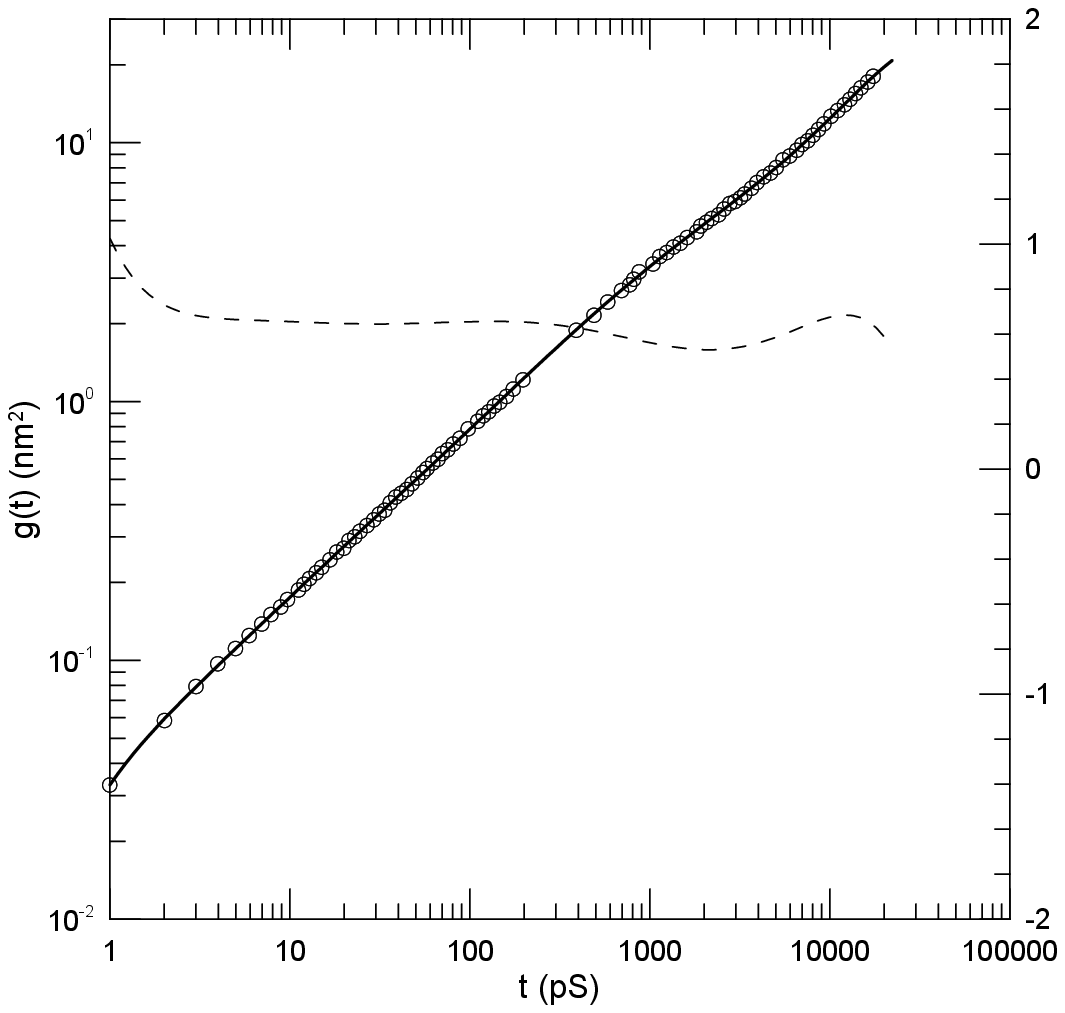}
  \caption{Mean-square atomic displacement $g(t)$ from a simulated polyethylene melt, results of Padding and Briels\cite{padding2001b}, their Figure 1.  Circles are digitized data from the original paper, the solid line is the polynomial fit to the data, and the dashed line is the first logarithmic derivative of the fitted function, showing that the fitted curve has a nearly uniform slope near 0.65, with a slight dip down to 0.53 at times between 600 and 5000 pS.}
  \label{figure:method1padding}
\end{figure}

The original authors report that $g(t)$ can be described by two power laws, one with $\alpha = 0.65$ for $t < 200$ pS and another with $t=0.57$ for $t > 200$ pS. On a log-log plot, a power law would appear as a straight line whose slope equal $\alpha$.   Our results are seen in Figure \ref{figure:method1padding}. As seen in the figure, for times of a few picoseconds up to 200 pS, the calculated slope is very nearly constant, corresponding to $\alpha \approx 0.65$ exactly as reported by Padding and Briels. For $t> 200$ pS our polynomial fit reveals that the slope decreases very slightly, to perhaps 0.53 or so, and then increases again back toward 0.65.  This gentle modulation of the slope is revealed by our fitting process, but would have been masked a fit to an assumed power-law behavior.  Such a power law is consistent with these measurements, as seen in Padding and Briels' Figure 2, the deviations of the data from a simple power law being modest.

The fitting process is seen to reveal power-law behavior when such is present.

Brodeck, et al.\cite{brodeck2010a} report atomistic molecular dynamics simulations of a polyethyleneoxide-polymethylmethacrylate mixture at four temperatures. Their interest was the dynamic asymmetry between the two components, with polymethylmethacrylate density fluctuations relaxing much more slowly than polyethyleneoxide density fluctuations. They report an analysis using Rouse mode decomposition, mean-square displacements, non-Gaussian parameters for the distribution of mean-square displacements, and comparison with a simple bead-spring model.

\begin{figure}[thb]
  \includegraphics{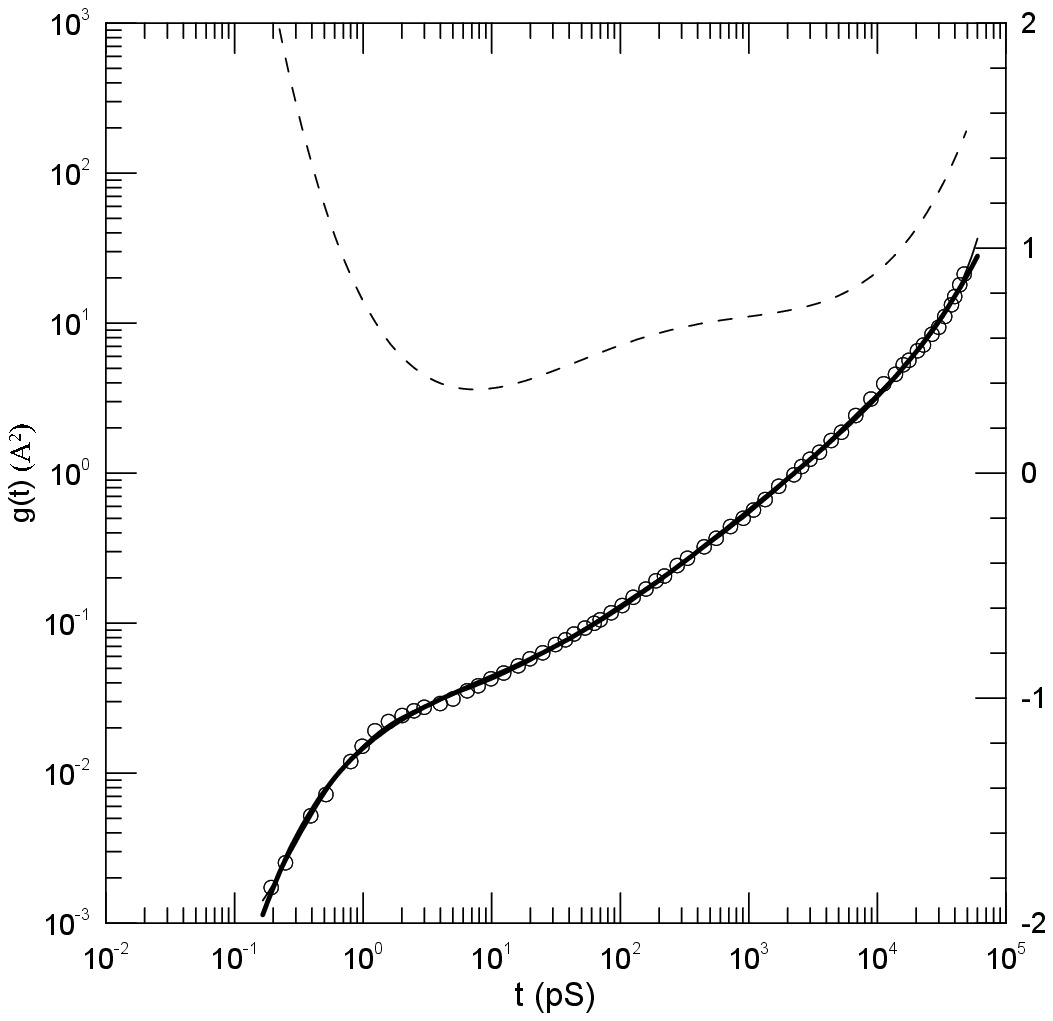}
  \caption{Determinations of the chain center-of-mass mean-square $g(t)$ from results of Brodeck, et al.\cite{brodeck2010a} at 400 K.  Circles are digitized data from the original paper, solid lines (indistinguishable over most of the graph) are polynomial fits of different order, and the dashed line is the first logarithmic derivative of the fitted curve, showing that the fitted curve has a minimum slope of 0.37 near $t=7$, the slope increasing uniformly toward a slope of 1.0 near $t=1\cdot 10^{4}$}
  \label{figure:methodtest1}
\end{figure}

Figure \ref{figure:methodtest1} shows their calculated mean-square center-of-mass displacements for their blend at 400 K.  Fits to fifth, sixth, and eighth order polynomials lead to the solid lines that pass very nearly uniformly through the data points.  The lines are almost completely overlapping, except note upper right.  The computed first logarithmic derivatives correspond to the dashed line (actually three lines, overlapping) at the top of Figure 2.  Over times $5-10^{4}$ pS, the slope smoothly increases from 0.37 to 1.0.  At the two extrema, the slope becomes considerably larger, reaching $\approx 2$ at short times.    The original authors displayed a $t^{1}$ tangent line at longer times $\sim 10^{4}$ pS). In that region, we find that $\log(g(t))$ follows a smooth curve having a continuously increasing slope, with, as reported by Brodeck, et al., a tangent having slope $\approx 1$ at times $\approx 1 \cdot 10^{4}$ pS.

The fitting process is seen to describe time dependences that are more complex than single power laws.

Peng, et al.\cite{peng2017a} report simulations of a united-atom model of flexible polymers having bead-bead Lennard-Jones interactions, beads of each chain being linked with a finitely extensible nonlinear elastic potential.  To these were added chains made more rigid by giving them significant bond-bending and torsional potentials. The blends were of interest because they contained two polymer species with very different glass temperatures and mobilities.

\begin{figure}[thb]
  \includegraphics{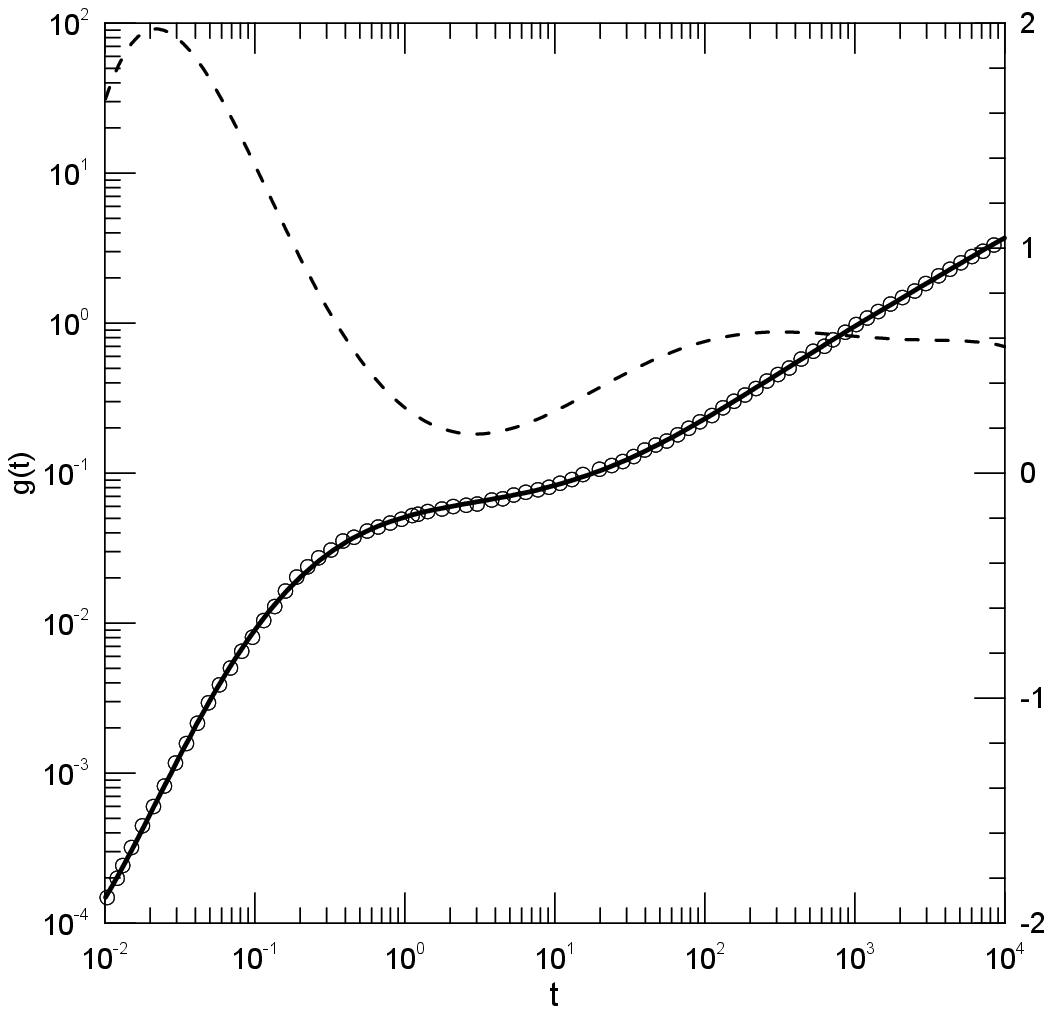}
  \caption{Mean-square atomic displacement $g(t)$ from  Peng, et al.\cite{peng2017a}.  Circles are digitized data from the original paper, the solid curve is the polynomial fit to the data, and the dashed line is the first logarithmic derivative of the fitted function.  The first derivative curve shows that the possible power law region near $t=3$ is actually an inflection point.}
  \label{figure:methodtestpeng}
\end{figure}

Figure \ref{figure:methodtestpeng} shows means-square displacements of beads of the non-rigid polymers in a mixture to which a small amount of the more rigid polymer has been added (Peng, et al.'s Figure 9a, $N_{A} = 5$).  The solid line represents an eighth-order polynomial fit, which describes well $\log(g(t))$  at almost all times.  We see for long times ($t >100$) that $\log(g(t))$ increases nearly linearly in time, with a slope $d \log(g(t))/d \log(t)  \approx 0.6 $. At short times ($t \leq 0.1$) $g(t)$ increases rapidly with increasing time, with a slope approaching 2.  Of particular interest for this paper is the behavior of $\log(g(t))$ for times near $t=3$.  In this range, $\log(g(t))$ superficially looks as though it increases linearly with increasing time, implying power-law behavior $g(t) \sim t^{\alpha}$.  However, the first derivative, the dashed line, simply has a minimum at $t=3$, the slope having a parabolic dependence on $\log(t)$ around this point.  The region near $t=3$ is thus revealed to be an inflection point, not a local power-law regime.

\section{Discussion}

Here we have demonstrated an approach to analysing simulated measurements of the mean-square displacement $g(t)$ of polymer molecules. The approach is of interest because there are theoretical models that predict that $g(t)$ is described by power laws in $t$, separated by transition regimes.  Power law regimes are revealed as regions of $t$ in which the logarithmic derivative of $g(t)$ is nearly constant.  As shown above, our approach finds such regions when they exist.  The same method also identifies regions whose time dependence is not a power law, and distinguishes between a power law regime and an inflection point in $\log(g(t))$.

This paper represents a proof of principle test.  It is appropriate to ask if the approach described here is generally applicable, as opposed to the above discussion reflecting the behavior of a few outlying results.  It is also appropriate to ask if the approach leads to new physically-interesting results. A full length paper, answering this question by analyzing close to a hundred sets of $g(t)$ data, is now in preparation.


\begin{thebibliography}{66.}

\bibitem{degennes1979a}  P.-G. deGennes.  \emph{Scaling Concepts in Polymer Physics}. Cornell U.P.: Ithaca (1979).

\bibitem{doiedwards1986a} M. Doi, and S. F. Edwards. \emph{The Theory of Polymer Dynamics}.  Oxford University Press, Oxford (1986).

\bibitem{padding2001b} J. T. Padding and W. J. Briels.  Zero Shear Stress Relaxation and Long Time Dynamics of a Linear Polethylene Melt: A Test of Rouse Theory.  \emph{J. Chem.\ Phys.} \textbf{114}, 8685-8693 (2001).

\bibitem{brodeck2010a}  M. Brodeck, F. Alvarez, A. J. Moreno, J. Colmennero, and D. Richter. Chain Motion in Nonentangled Dynamically Asmmetric polymer Blends: Comparison between Atomistic Simulations of PEO/PMMA and a Generic Bead-Spring Model. \emph{Macromolecules}\textbf{43}, 3036-3051 (2010).

\bibitem{peng2017a} W. Peng, R. Ranganathan, P. Keblinski, and R. Ozisik.  Viscoelastic and Dynamic Properties of Well-Mixed and Phase-Separated Binary Polymer Blends: A Molecular Dynamics Simulation Study. \emph{Macromolecules}  \textbf{50} 6293-6302 (2017).

\bibitem{padding2001c} J. T. Padding and W. J. Briels.  Uncrossability Constraints in Mesoscopic Polymer Melt Simulations: Non-Rouse Behavior of C$_{120}$H$_{242}$.  \emph{J. Chem.\ Phys.} \textbf{115}, 2846-2859 (2008).

\bibitem{padding2002a} J. T. Padding and W. J. Briels. Time and Length Scales of Polymer Melts Studied by Coarse-Grained Molecular Dynamics Simulations.   \emph{J. Chem.\ Phys.}  \textbf{117}, 925-943 (2002).

\bibitem{padding2003b} J. T. Padding and W. J. Briels. Coarse-Grained Molecular Dynamics Simulations of Polymer Melts in Transient and Steady Shear Flow. \emph{J. Chem.\ Phys.} \textbf{118}, 10276-10286 (2003).


\end{thebibliography}
\end{document}